# Phosphorene: Fabrication, Properties and Applications

Liangzhi Kou [†*], Changfeng Chen[‡] and Sean C. Smith [†*]

[†] Integrated Materials Design Centre (IMDC), School of Chemical Engineering, UNSW Australia, Sydney, NSW 2052, Australia.

[‡] Department of Physics and Astronomy and High Pressure Science and Engineering Center, University of Nevada, Las Vegas, Nevada 89154, United States

**Abstract**

Phosphorene, the single- or few-layer form of black phosphorus, was recently rediscovered as a two-dimensional layered material holding great promise for applications in electronics and optoelectronics. Research into its fundamental properties and device applications has since seen exponential growth. In this Perspective, we review recent progress in phosphorene research, touching upon topics on fabrication, properties, and applications; we also discuss challenges and future research directions. We highlight the intrinsically anisotropic electronic, transport, optoelectronic, thermoelectric, and mechanical properties of phosphorene resulting from its puckered structure in contrast to those of graphene and transition-metal dichalcogenides. The facile fabrication and novel properties of phosphorene have inspired design and demonstration of new nanodevices; however, further progress hinges on resolutions to technical obstructions like surface degradation effects and non-scalable fabrication techniques. We also briefly describe the latest developments of more sophisticated design concepts and implementation schemes that address some of the challenges in phosphorene research. It is expected that this fascinating material will continue to offer tremendous opportunities for research and development for the foreseeable future.

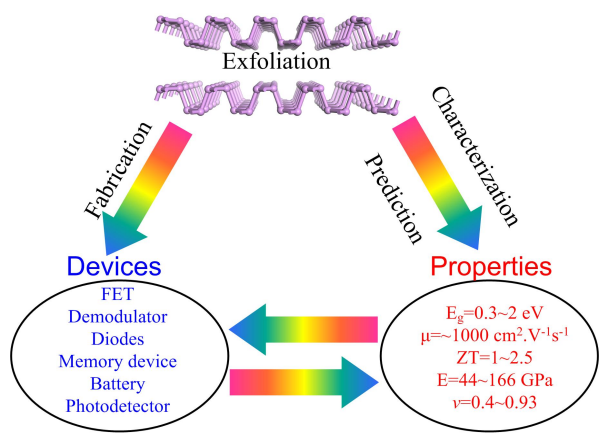

Email: kouliangzhi@gmail.com, sean.smith@unsw.edu.au

Phosphorus is a common material with industrialized production, and it has been extensively used in matchsticks, fireworks, chemical fertilizers, and the napalm bomb[1]. Since its discovery in 1914 this material has not attracted any special attention over the past century by physicists and chemists due to its structural instability and strong toxicity[1]. In recent years, there has been tremendous interest in two-dimensional (2D) layered materials[2], following the discovery of graphene[3] and the transition-metal dichalcogenide (TMDC) family[4,5]. The quest for a variety of high-performance devices has necessitated the search for additional layered materials that exhibit a wider operating range in their key properties, such as the electronic band gap and carrier mobility. Possessing a stacked layered structure and weak van der Waals (vdWs) interlayer interactions[6], black phosphorus (BP) was introduced as a new member of the 2D layered material family; it is the most stable allotrope amongst the group also including white, red and violet phosphorus.[7,8] The successful fabrication and outstanding performance of the field-effect transistors (FETs) constructed using the exfoliated BP by two independent groups[9,10] in early 2014 ignited a surge of research activities in the physics, chemistry and materials communities. Extensive theoretical and experimental investigations in the past year or so have produced findings of novel physical, chemical and mechanical properties, and led to the fabrication of phosphorus-based nanodevices with promising potentials for applications in electronics, optoelectronics, photovoltaics and spintronics[11,12].

The novel properties of BP are fundamentally rooted in its unique structural arrangement, which has a strong in-plane covalent bonding network but weak vdWs interlayer interactions. Its interlayer distance between 3.21 to 3.73 Å depending on the stacking pattern[6,13] makes it highly suitable for mechanical exfoliation as in the production of graphene[14] and TMDCs[15-16]. Monolayer or few-layer black phosphorus, also known as phosphorene, has been successfully obtained using the cleavage approach[9,10], and it joins the 2D layered material family that also includes graphene[3], graphene analogues (like silicene, gemanene, BN and so on)[2,17] and TMDCs (with formula of MX$_2$; M=Mo, W, Nb and Ta; X=S, Se and Te)[4,5]. Phosphorene distinguishes itself from other 2D layered materials by its unique structural characteristics: it has a puckered structure along the armchair direction (Fig. 1a), but it appears as a bilayer configuration along the zigzag direction (Fig. 1b). This structural anisotropy can be clearly seen in its local bonding configurations. The bond angle along the zigzag direction, known as the hinge angle, is 94.3° and the adjacent P-P bond length is 2.253 Å; these values are smaller than the corresponding dihedral angle along the zigzag direction (103.3°) and the connecting bond length (2.287 Å). The lattice constants along the two perpendicular directions are different, at 3.30 Å and 4.53 Å, respectively.[10] Such a unique structural arrangement, resembling a network of connected hinges (see Fig. 1c, 1d), is the origin of the anisotropic physical (electronic band structure[10,18-19], electrical transport[10, 20], thermoelectric[21-24]) and mechanical (critical strain[25], Poisson's ratio[26-28], Young's modulus[25,26,28]) properties, and it has led to distinct nanodevice designs exploiting this directional selectivity[29-31].

Since the successful fabrication of phosphorene-based FET[9-10] in early 2014, there has been an explosive growth of research that has unveiled many new properties with great potential for novel applications. These new findings cover a wide range of topics, including intrinsic electronic and transport properties[10,18-20], engineering of electronic properties by strain[18-19], electric field[13] or transition metal doping and small molecule adsorption [32-35], low-dimensional phosphorene derivatives[36-40] and new phosphorene polymorphs[41-43], novel physical phenomena (e. g., topological insulators[44], superconductivity[45], and even the integer quantum Hall effect[46]), unique mechanical features (e. g., negative Poisson's ratio and anisotropic Young's modulus)[25-28], and promising device applications in electronics[47-48], photovoltaic[49-52] and gas sensors[30]. Despite these extensive reported studies, investigations into the fundamental properties and applications of phosphorene are still at an early stage. It is probably fair to say that research on 'simple phosphorene' has not reached its zenith, with new concepts and properties being unveiled constantly. However, since most of the 'low-hanging phosphorene fruits' have already been harvested, many research activities are now shifting from studying phosphorene itself to the use of the material in various applications and as a versatile platform for further investigations of more complicated phenomena. The exploration of the fundamental science and technological development associated with phosphorene remains far from being exhausted, and as the quality of phosphorene material and devices continues to improve, more breakthroughs can be expected.

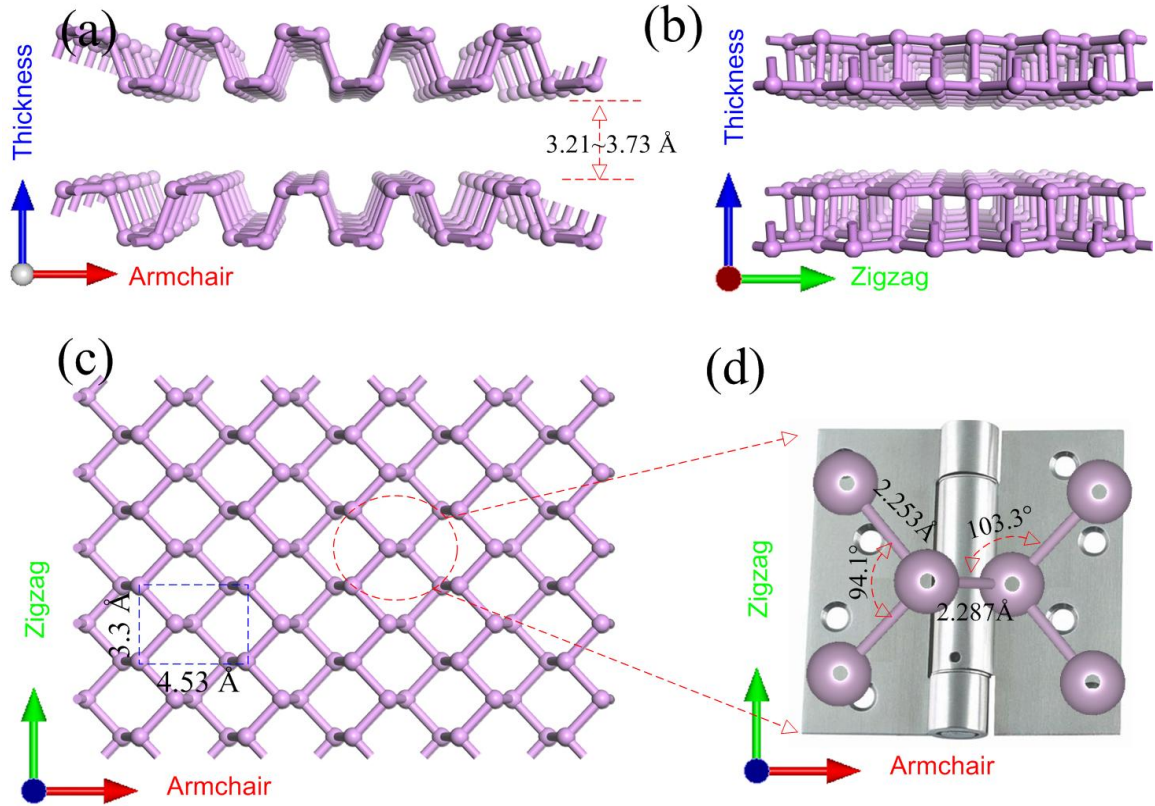

Figure 1. Structure of phosphorene: (a,b) side views from the zigzag and the armchair direction, (c) top view, and (d) zoomed-in local atomic structure of the P-P bonding configuration, where each phosphorus atom is bonded with three adjacent phosphorus atoms, forming a puckered structure, which resembles a network of connected hinges that exhibits anisotropic rotational deformation under strain[31]. The different bonding arrangements along the armchair and zigzag direction are the origin of the intrinsic anisotropy of many physical and mechanical properties of phosphorene.

In this Perspective, we review recent experimental and theoretical progress in phosphorene fabrication, properties and applications, and discuss challenges and possible future research directions. Fabrication of phosphorene is currently achieved mainly by mechanical exfoliation, which is limited to laboratory usage; advanced growth strategies for large-scale production are required for practical applications. The unique anisotropic properties of phosphorene are highlighted in terms of its electronic structure, optical and electrical conductance, thermoelectric performance, topological phase, as well as unusual mechanical behaviour such as the negative Poisson's ratio. We emphasize that computational studies play a vital role in helping accelerate scientific discovery by facilitating the interpretation of important experimental findings and offering insights for technological advancement of new material and device design. Since the successful fabrication of few-layer BP and discovery of its novel properties, demonstrations of its viability in various device settings have been reported, and there are also a wide range of predictions and proposals for its potential implementation in a variety of innovative applications. However, we bring attention to issues like the vulnerability of phosphorene to degradation by oxidation and other surface reactions that have to be overcome before practical usage can be realized beyond the stringent conditions in the laboratory. Finally, we provide an outlook about possible future work to further explore fundamental properties of phosphorene and optimize the performance of phosphorene-based materials and devices.

## Experimental Fabrication of Phosphorene

Reliable production of atomically thin, layered phosphorene with uniform size and properties is essential to translating their superior properties into high-performance device applications. Here we review and comment on current methods for fabrication and evaluate their merits.

**Top-down methods:** The isolation of single-layer graphene by mechanical exfoliation has unleashed a new research field devoted to the study of the properties of 2D materials.[14-16] This method has proved to be effective to cleave bulk layered materials down to the single- and few-layer limits. One important precondition for mechanical exfoliation is that the interlayer interaction in the bulk counterpart is dominated by weak vdWs forces, making possible the cleavage of the materials using just an adhesive tape. The structural characteristic of BP, where its layered structure is held together by weak interlayer forces with a significant vdWs character, suggests that few-layer or even monolayer phosphorus can be obtained by the exfoliation method. The energy required to exfoliate layered crystals can be quantified by the surface energy[5], which is the energy to remove a monolayer from the crystal divided by twice the monolayer surface area. Although no measured data on BP is currently available, the theoretically predicted binding energy[53] is similar to those experimentally determined for BN, $MoS_2$, $WS_2$ and $MoSe_2$ (65-75 mJ m$^{-2}$)[54] or graphene (65-120 mJ m$^{-2}$)[55-56].

Atomically thin flakes of phosphorene can be peeled from their parent bulk crystal (BP), which is commercially available, by micromechanical cleavage using an adhesive tape[9-10]. Following transfer onto the $Si/SiO_2$ substrates, see Fig. 2a, the band gap of phosphorene is measured by photoluminescence spectra to be 1.45 eV. All samples were sequentially cleaned after the transfer by acetone, methanol, and isopropyl alcohol to remove any scotch tape residue. This procedure has been followed by a 180 °C post-bake process to remove the solvent residue. Other layered materials such as BN[57] can also be mechanically exfoliated into single sheets by this method. Mechanical cleavage produces single-crystal flakes of high purity and cleanliness that are suitable for fundamental characterization[9-10] and for fabrication of individual devices[47-52]. However, this method is not scalable and hence is limited to be used in the laboratory; another issue is a lack of systematic control of flake thickness and size.

Meanwhile, some improved fabrication methods have been suggested. One example is the plasma assisted process as also used for graphene and $MoS_2$.[58-59] Recently, Lu et al.[60] demonstrated that a focused Ar+ plasma (commercial 13.56 MHz RF source with a power of 30 W and a pressure of 30 Pa) can be used to thin few layer phosphorene down to monolayer thickness by thermal ablation with micrometre-scale resolution, which can be identified by optical contrast spectra and AFM, see Fig. 2b. This method combined with exfoliation provides an improved way for controlling thickness of phosphorene, but the requirement for laser raster scanning makes it challenging for scale-up applications. An alternative approach, such as the modified mechanical exfoliation method[61] that employs a silicone-based transfer layer to optimize the yield of atomically thin BP flakes, can improve the samples to some degree, but this is still only suitable for small-scale production for fundamental research.

To obtain larger quantities and sizes of exfoliated phosphorene nanosheets, liquid-phase preparation techniques are very promising[62-63]. The interlayer intercalation of ionic species in BP expands the interlayer distance to allow the layers to be exfoliated in liquid. This approach has been successfully applied for fabrication of TMDC, hexagonal BN, bismuth telluride ($Bi_2Te_3$) and graphene[63], which have shown good performances in devices. It is thus expected that solution-based phosphorene production will also have similarly good prospects for making flexible electronics[64-65]. Typically, a chunk of black phosphorous crystal is immersed into different solvents (such as alcohols, chloro-organic solvents, ketones, cyclic or aliphatic pyrrolidones, N -alkyl-substituted amides, and organosulfur compounds) to allow the ionic species to intercalate. The sample is sonicated to break down the interlayer vdWs bonding and accelerate the process. Yasaei et al.[64] found that aprotic and polar solvents such as dimethylformamide (DMF) and dimethyl sulfoxide (DMSO) are appropriate solvents for the synthesis of atomically thin phosphorene nanoflakes and can produce uniform and stable dispersions after sonication. The solutions were then centrifuged and their supernatants were carefully collected by a pipette. The exfoliated nanoflakes (Fig. 2c) show competitive electrical properties compared to mechanically exfoliated BP flakes. This opens up new possibilities for the atomically thin BP layers to be formed into thin films and composites on large scales with a wide range of applications such as flexible electronics and optoelectronics.

Another promising alternative method that is faster and more controllable is lithiation where an electrochemical cell with lithium foil anode and layered-material-containing cathode are used[66]. The resulting Li-intercalated material is exfoliated by sonication in water like liquid phase preparation as mentioned above, yielding monolayers. This method has been demonstrated for $MoS_2$, $WS_2$, BN and graphene,[66-67] but its use for phosphorene fabrication has not yet been reported. One possible problem of the method for phosphorene fabrication is that the element of lithium may strongly interact with phosphorene and not easy to move. Recent calculations indicate that lithiation will cause only a small (0.2%) volume change in phosphorene[68], and the interlayer intercalation will weaken the vdWs interaction, leading to exfoliation with the help of ultrasonication[63] as seen in graphene and TMDC materials.

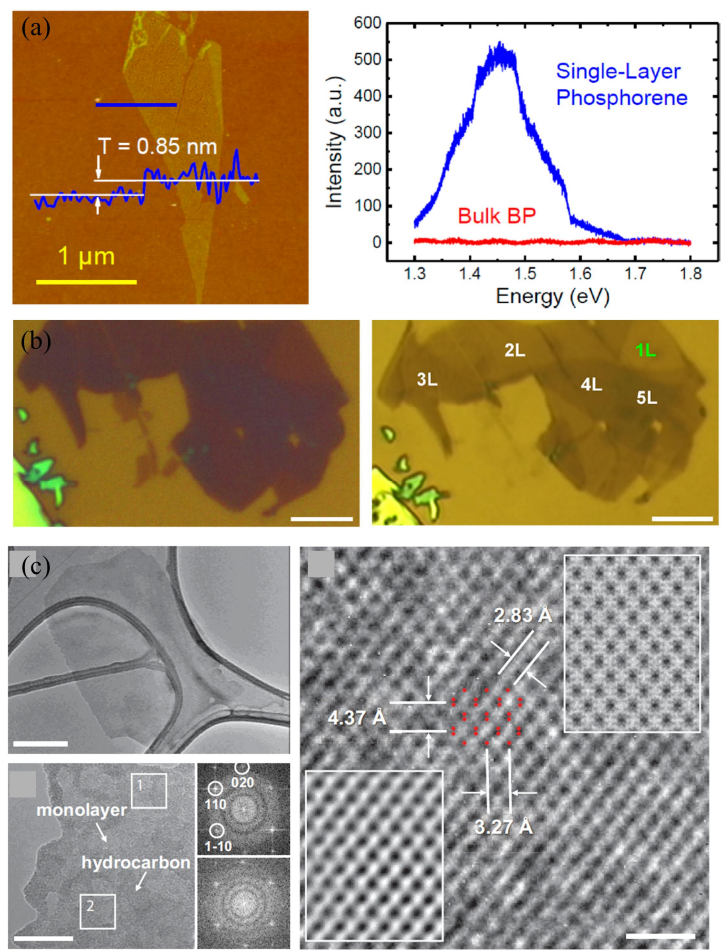

Figure 2 Synthesis of phosphorene. (a) Atomic force microscopy (AFM) image of a single-layer phosphorene crystal with the measured thickness of ≈0.85 nm from exfoliation method, and photoluminescence spectra for single-layer phosphorene and bulk BP samples on a 300 nm $SiO_2$/Si substrate, showing a pronounced PL signal around 1.45 eV. (b) Optical image of multilayered pristine phosphorene before and after Ar+ plasma thinning. The scale bar is 5μm. (c) Top left: a typical TEM image of a BP flake on lacy carbon support (scale bar is 200 nm) from liquid exfoliation. Bottom left: a TEM image and the FFT of the selected area, suggesting uniform

existence of single layer BP over the entire flake. Right: the corresponding high resolution TEM image of monolayer BP nanoflake. **a: Reprinted from Ref. 10; b: Reproduced from Ref. 40 with permission from Springer; c: Reproduced from Ref. 64 with permission from Wiley.**

**Bottom-up methods:** The development of wafer-scale synthesis methods via chemical vapour deposition (CVD) on metal substrates[69-70] and epitaxial growth on insulating substrates[71-72] has enabled large-scale device fabrication based on graphene and TMDCs. Compared with mechanical exfoliation, investigations of chemical synthesis of large-area and uniform phosphorene are rare. The recent growth of single crystals of orthorhombic BP from red phosphorus and Sn/SnI$_4$ as mineralization additive provide a new possibility.[73] Sample sizes of several millimetres can be realized with high crystal quality and purity, making it possible for large-area preparation of single- or multi-layer phosphorene. Direct chemical growth strategies for phosphorene are still lacking at present probably due to its chemically active surface that is fragile when exposed to air, as well as the absence of a suitable substrate for its CVD growth; however, the successfully synthesized monolayer silicene, germanene and stanene with strong chemically active surface on the substrates (for instance silicene grown on Ag substrate)[74] provided good examples for possible phosphorene chemical growth; its development and implementation will require concerted efforts from the materials and chemistry communities.

Table 1 summarizes the recent progress of phosphorene fabrication in comparison with the more mature and well-studied graphene and TMDCs. It is clear that the top-down method (mechanical exfoliation and related improved methods), which has been extensively used for graphene and TMDCs, is successful in phosphorene fabrication except lithiation. This means that the small-scale fabrication of monolayer or few-layer phosphorene is relatively mature, which can be used for device demonstrations, such as FETs and optoelectronics.[9-10,47-52] In contrast, the bottom-up approach such as chemical synthesis is almost totally blank for phosphorene. This inhibits large-area and uniform phosphorene growth and its practical applications; it also highlights a large unexplored area for chemical synthesis of phosphorene. Advanced chemical synthesis of phosphorene may draw inspiration from the techniques used for graphene and TMDCs, such as hydrothermal synthesis[75-76] or CVD growth[69-72].

**Table 1 Summary of fabrication methods for phosphorene in comparison with those for graphene and TMDCs.**

| | Phosphorene | | Graphene | | TMDC | |
|---|---|---|---|---|---|---|
| **Top-down** | | Ref. | | Ref. | | Ref. |
| Cleavage with Tape | √ | 9, 10 | √ | 14 | √ | 14, 16 |
| Liquid-phase exfoliation | √ | 64, 65 | √ | 56, 62 | √ | 62, 63 |
| Lithiation | × | | √ | 66 | √ | 66, 67 |
| Plasma-assisted fabrication | √ | 60 | √ | 58 | √ | 59 |
| **Bottom-up** | | | | | | |
| CVD growth | × | | √ | 69, 72 | √ | 70, 71 |
| Hydrothermal synthesis | × | | √ | 75 | √ | 76 |

## Physical and mechanical properties of phosphorene

Phosphorene has been shown to possess many intriguing properties originating from its unique structure (see Fig. 1). In this respect, theoretical predictions are advancing ahead of experimental measurements. Computational studies can help interpret experimental findings and, more importantly, accelerate scientific discovery and technological advances through simulations of material behaviour under diverse conditions, which may offer crucial insights for new device concept and design. As more experimental data become available, the interplay between theory and experiment will prove to be even more powerful. Below we review the current status on the understanding of physical and mechanical properties of phosphorene with an emphasis on its unique intrinsic anisotropy in structure and property. Comparative notes are also made between phosphorene, graphene and TMDCs.

**Electronic structure**

Bulk BP is a semiconductor with a direct band gap of 0.3 eV determined by first-principles calculations ($G_0W_0$)[77] and angle-resolved photoemission spectroscopy (APERS)[9, 48]. As the thickness decreases, the band gap gradually increases due to quantum confinement, reaching 2 eV for monolayer phosphorene, following the formula $A/N^{0.7}+B$ (N is the layer number), which decays significantly slower than the usual $1/N^2$ scaling predicted by the standard quantum confinement effect. Unlike the indirect-to-direct band gap transition in TMDCs, the band topology remains the same with changing thickness; all few-layer phosphorene samples are direct semiconductors with the conduction band minimum at the $\Gamma$ point (Fig. 3a), although the valance band maximum is slightly away from Γ (by only 0.06 2π/a$_y$)[18]. The thickness independent band topology of phosphorene is important for its potential photonics and optoelectronics applications. The gap value (0.3-2 eV) is typically smaller than those of the TMDC compounds (1.1-2.5 eV), but larger than the semi-metallic graphene, enabling phosphorene to possess a moderate on/off ratio ($10^4$-$10^5$)[9-10] while preserving a sufficiently large carrier mobility (around 1000 cm$^2$/V.S)[9,48] suitable for many applications. Phosphorene thus has considerable advantages in its potential uses in semiconductor devices. In comparison, graphene has extremely high mobility, but its lack of an intrinsic band gap renders it impractical in FET applications due to a small on/off ratio. Meanwhile, MoS$_2$ has a large band gap and high on/off ratio ($10^8$), but the much smaller carrier mobility (200 cm$^2$/V/s in monolayer MoS$_2$[5]; note that recent studies show that even the mobility of monolayer MoS$_2$ is overestimated[78-79]) limits its widespread application in electronics.

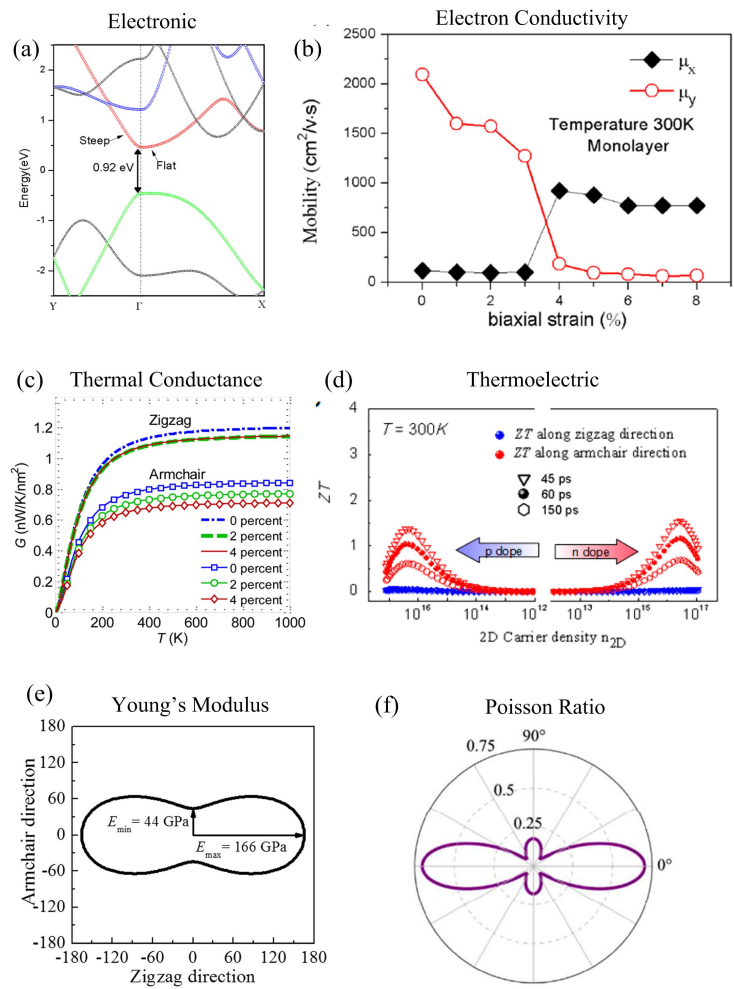

Figure 3. Intrinsic anisotropy of physical and mechanical properties of phosphorene. (a) Electronic band structure of monolayer phosphorene, where highly anisotropic band dispersions along the Γ-x and Γ-y directions are clearly visible. (b) Electron mobility of monolayer phosphorene at room temperature and its modulation by strain. (c) Anisotropic thermal conductance along the armchair and zigzag directions at different temperatures. (d) Thermoelectric figure of merit, ZT, as a function of doping carrier density at 300 K. (e) Young's modulus showing obvious directional anisotropy. (f) Poisson's ratio showing strong directional anisotropy. **a, b: Reprinted from Ref. 19; c, d: Reprinted from Ref. 24; e: Reproduced from Ref. 25 with permission from American Institute of Physics; f: Reproduced from Ref. 26 with permission from Royal Society of Chemistry.**

Originating from its unique structural characteristics (see Fig. 1), the most impressive feature of phosphorene is its highly anisotropic band dispersion[19] in the Brillouin zone (BZ) near the Fermi level as shown in Fig. 3a. Both the top of the valence bands and the bottom of the conduction bands show much more significant dispersions along the armchair direction than in the zigzag direction where these bands are nearly flat. Therefore, the corresponding effective mass of electrons and holes are also highly anisotropic since it is proportional to the inverse of the curvature of the band dispersion. This anisotropic effective mass or band dispersion is responsible for the recently observed anisotropic electrical conductance and electron mobility (Fig. 3b) in phosphorene[10], where the value along the armchair direction is three orders of magnitude larger than that along the zigzag direction. It is interesting to note that the direction of the electrical conductance can be rotated 90° under a biaxial or uniaxial strain of 4-6%, resulting from a switch in the energetic order of the first and second lowest-energy conduction bands induced by strain (red and blue lines in Fig. 3a). It leads to a significant enhancement of thermoelectric figure of merit in phosphorene.

The electronic properties of phosphorene are highly sensitive to external field and doping. The gap value depends on the interlayer stacking pattern; for instance, the band gap of the bilayers can vary from 0.78 eV to 1.04 eV in different stacking orders due to the effect of interlayer interactions[13], where the interlayer distance is in the range of 3.21-3.73 Å. This wide range of tunability of the band gap has a profound influence on other properties. Strain engineering is an effective and economical approach to modifying and controlling electronic properties of nanomaterials. First-principle calculations reveal that monolayer phosphorene can withstand a tensile stress and strain up to 10 N/m and 30%, respectively. The band gap experiences a direct-indirect-direct transition when axial strain is applied[80]. Calculations also predict a semiconductor-to-metal transition in bilayer phosphorene under a normal compressive strain[81]; however, 2D materials tend to crumple under compressive deformation, which casts doubt on the simulations that omitted such a possibility. The band gap is shown to be robust against compressive strains along the zigzag direction due to ripple deformation.[31] Besides the strain effect, electric field has been shown to modulate the band gap of phosphorene[13,36], even inducing a topological phase transition (with a 5 meV nontrivial gap).[44] Effective modulation and control of the electronic properties of phosphorene can also be achieved through cutting 2D phosphorene into 1D ribbons, edge doping, and coupled external strain/electric field approaches[36-39]. Furthermore, elemental and functional group decoration of phosphorene has been proposed to tune the electronic properties for tailored applications[33-35]. For instance, the introduction of 3$d$ transition-metal atoms induces magnetism in phosphorene, where the magnitude of the magnetic moment depends on the metal species, and the results can be tuned by the applied strain[33, 35].

**Electrical/thermal conductivity and thermoelectric performance**

Phosphorene displays promising potential for thermoelectric applications. Thermoelectric devices rely on the Seebeck effect to convert heat flow into electricity, and the Seebeck coefficient is proportional to the ratio of a device's electrical conductance to its thermal conductance. The overall device efficiency is measured by the thermoelectric figure of merit, ZT. To maximize the ZT value, it is desirable to achieve simultaneously a high electrical conductance and a low thermal conductance, which presents a formidable challenge since the electrical and thermal transport are positively correlated in most materials. The intrinsic anisotropy of lattice and electronic properties in phosphorene offers a unique solution to this challenge. The prominent electron transport in phosphorene occurs along its armchair direction, which coincides with the direction for poor thermal conductance, see Fig. 3c,[24] where the predicted thermal conductivity is 36 W/m-K, which is only about one third the value (110 W/m-K) along its zigzag direction[23]. Such a favourable alignment of good electrical conductance and poor thermal conductance would significantly enhance phosphorene's thermoelectric performance. From first-principles calculations,[24] the ZT value in monolayer phosphorene can reach up to 2.5 along the armchair direction at 500 K and doping density of $2 \times 10^{16}$ m$^{-2}$. At such a doping density, ZT value is still larger than 1.0 at room temperature, 300 K (Fig. 3d). From these results, phosphorene-based thermoelectric devices can reach an energy conversion efficiency of 15-20%, meeting the criterion for commercial use. Hence, phosphorene is an outstanding candidate material for thermoelectric devices. It is noted that from simulation results, both the thermal and electrical conductance can be modulated by strain and optimal doping level;[21-24] it is thus expected that the thermoelectric performance (ZT value) can also be engineered by strain or doping like those for electronic modulation.

The investigations of thermoelectric properties illustrate the importance of structural and property anisotropy of phosphorene in producing a superior level of device performance. It is instructive to make a comparison with graphene and $MoS_2$. Graphene shows a maximum Seebeck coefficient as large as 30 mV/K[82], but its very high thermal conductivity of 2000-5000 W/mK renders the ZT value of graphene impractically small. Although such high thermal conductivity can be suppressed in graphene ribbons or by introducing disorder, the possibility of graphene's thermoelectric application is still a subject of debate.[83] The thermal conductivity of $MoS_2$ is estimated to be 52 W/mK[84], which is comparable with the corresponding value of phosphorene, but the ZT value is generally lower than 0.4 at room temperature[85]. It is clear that the overall thermoelectric performance of phosphorene is superior, indicating significant promise for phosphorene in thermoelectric devices.

**Mechanical properties**

The structural anisotropy of phosphorene suggests significantly anisotropic mechanical response to uniaxial loading along the armchair and zigzag directions. From theoretical calculations, a monolayer phosphorene can sustain tensile strains up to 27% and 30% in the zigzag and armchair directions, respectively[25]. There is, however, a pronounced difference in the deformation modes under different strain conditions. While tensile strain in the zigzag direction induces a larger P-P bond elongation, tensile strain applied in the armchair direction stretches the pucker of phosphorene without significantly extending the P-P bond length. Such characteristics lead to a direction-dependent Young's modulus (166 GPa along the zigzag direction and 44 GPa along the armchair direction), see Fig. 3e. Phosphorene demonstrates superior flexibility while its Young's modulus is considerably smaller than those of graphene (1 TPa) or $MoS_2$ (270 Gpa), see Table II. This property is especially useful in practical large-magnitude strain engineering of a variety of properties of phosphorene.

Phosphorene also exhibits an anisotropic Poisson's ratio, and its value along the zigzag direction (0.62 from Ref. 25; 0.93 from Ref. 27) is 2-4 times larger than that in the armchair direction (0.17 or 0.4), see Fig. 3f. It is expected to generate anisotropic out-of-plane structural fluctuation under compressive strains, and such anisotropic ripple deformation in phosphorene has been confirmed by first-principles calculations and the derived analytical expression from the classical elasticity theory along arbitrary strain directions[31]. It is shown that phosphorene only develops ripple deformation along the zigzag direction, which stems from its puckered structure with coupled hinge-like bonding configurations and the resulting anisotropic Poisson's ratio, where the out-of-plane deformation effectively releases the strain energy. In stark contrast, compression induced deformation along the armchair direction is dominated by bond-angle distortion without any appreciable ripple formation[31].

**Additional properties and comparison with other 2D materials**

Remarkably, phosphorene has been predicted to become superconducting with an estimated transition temperature of 3-16 K driven by doping and biaxial strain[45]; it is also suggested to be a superior gas sensor due to its adsorption sensitive surface and direction-selective I-V response[30]. The numerous structurally isomeric forms, such as blue phosphorene[42], phosphorene nanotubes and nanoribbons[36] further broaden the range of the phosphorene family and enrich their physical, chemical and mechanical properties.

Table 2 summarizes the electronic and mechanical properties of phosphorene and makes comparison with the corresponding parameters of graphene and TMDCs (with $MoS_2$ as a representative example). It can be clearly seen that phosphorene exhibits a comprehensive range of desirable physical parameters for device applications, making it a more favourable choice on balance than graphene and $MoS_2$, which show superior properties in some regards but lack the necessary characteristics in others. This overall picture bodes well for phosphorene as an outstanding member among the growing family of 2D layered materials, which serves as a strong motivation for further research and development of phosphorene to explore its applications in electronics, optoelectronics and photovoltaics.

**Table 2. Summary of electronic and mechanical properties of monolayer phosphorene compared to those of graphene and $MoS_2$. The values in the brackets are those along the zigzag direction in contrast to the preceding value along the armchair direction.**

|  | Phosphorene | Graphene | $MoS_2$ |
|---|---|---|---|
| **Band Gap** | 0.3-2 eV [9] | 0 [3] | 1.2-1.8 eV [5] |
| **Effective Mass** | 0.146 $m_e$ (1.246) [19] | ~0 [3] | 0.47-0.6 $m_e$ [86] |
| **Carrier Mobility** | ~1000 $cm^2.V^{-1}.s^{-1}$ [9] | 200,000 $cm^2.V^{-1}.s^{-1}$ [87] | 200 $cm^2.V^{-1}.s^{-1}$ [5] |
| **On/Off Ratio** | $10^3$-$10^5$ [9-10] | ~5.5-44 [88] | $10^6$-$10^8$ [5] |
| **Thermal Conductance** | 36 (110) W/m-K [23] | 2000-5000 W/m-K [83] | 52 W/m-K [84] |
| **ZT** | 1-2.5 [24] | ~0 [83] | 0.4 [85] |
| **Critical Strain** | 27% (30%) [25] | 19.4-34% [89] | 19.5-36% [90] |
| **Young's Modulus** | 44 (166) GPa [25] | 1 TPa [89] | 270$\pm$100 GPa [91] |
| **Poisson's Ratio** | 0.4 (0.93) [27] | 0.186 [89] | 0.21-0.27 [90] |

# Device applications

The easy, albeit small-scale, fabrication and novel properties of phosphorene have led to the design and demonstration of extensive device applications in the laboratory in the areas of electronics, batteries, optoelectronics, photovoltaics and so on. The extremely high pace of early developments suggests remarkable potential and many more novel devices for phosphorene applications. Here, we highlight two representative areas of device applications of phosphorene in transistors and batteries, and discuss other topics of possible future work.

**Transistor applications** One of the most important applications of semiconductors is for transistors in digital electronics. Recent decades have seen the continuous size reduction of transistors, currently down to about 20 nm, but this length scale is approaching statistical and quantum limits and also encountering difficulties associated with heat dissipation. This situation has presented a strong motivation to search for new device concepts and materials.

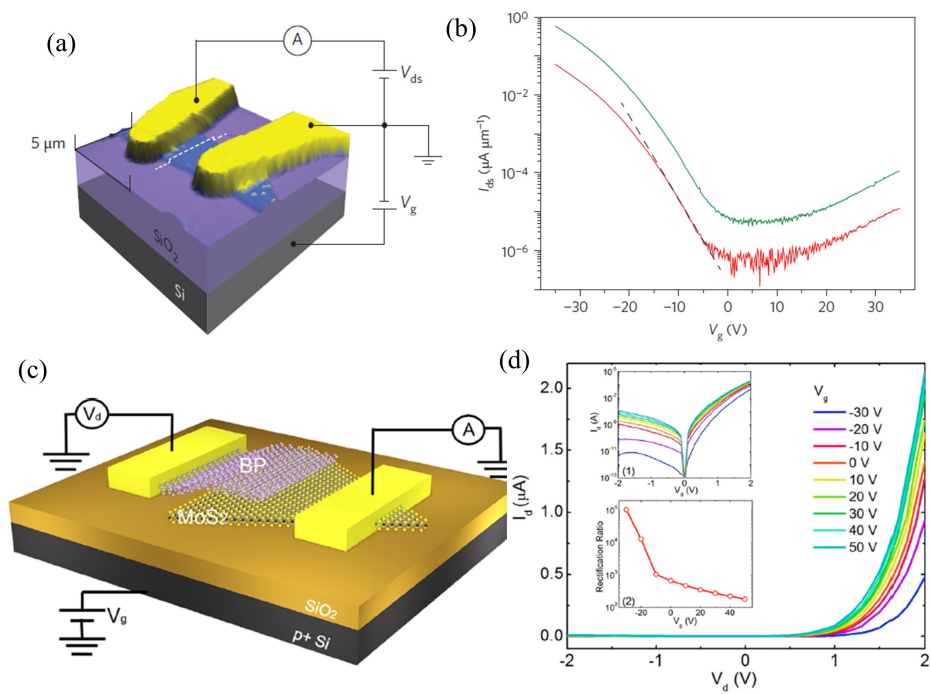

Figure 4. (a) Schematic device structure of phosphorene FET, where metal contacts were deposited on black phosphorus thin flakes by sequential electron-beam evaporation of chromium and gold through a stencil mask aligned with the sample. (b) Source–drain current as a function of gate voltage obtained from a 5-nm-thick device on a silicon substrate with 90 nm $SiO_2$ at room temperature, with drain–source voltages of 10 mV (red curve) and 100 mV (green curve). Drain current modulation up to $10^5$ is observed for both drain–source biases on the hole side of the gate doping. (c) Schematics of the device structure. Few-layer phosphorene flakes were exfoliated onto monolayer $MoS_2$ in order to form a vdWs heterojunction, while Ni and Au were deposited as contacts. (d) Gate tunable I-V characteristics of the 2D p-n diode. The current increases as the back gate voltage increases. <span style="color:red">The inset above shows the I-V characteristics under a semilog scale. The inset below shows the</span>

Desirable candidate materials for digital transistor applications should possess a high charge carrier mobility for fast operation, a high on/off ratio (requiring a moderate band gap larger than 0.4 eV)[5] for effective switching, high conductivity and low off-state conductance for low power consumption. All these factors are indispensable for successful operations. Graphene possesses exceptionally high carrier mobility (200,000 cm$^2$V$^{-1}$s$^{-1}$)[87], but its semi-metallic nature means that it cannot achieve a low off-state current or a high on/off ratio (typically around 5)[88], which limits its use as a digital logic device. TMDC layers generally have a band gap around 1.5 eV, and therefore have a very high on/off ratio ($1 \times 10^8$ for single layer MoS$_2$) and ultralow standby power dissipation; however, the moderate carrier mobility (200 cm$^2$ V$^{-1}$s$^{-1}$, or even lower) limits their performance. 2D materials with a more comprehensive set of desirable properties are highly sought after, and phosphorene seems to meet almost all the criteria for a good transistor material, see Table 2. Its tunable band gap of 0.3-2 eV ensures a large on/off ratio for phosphorene-based transistors, and its mobility of 600 cm$^2$V$^{-1}$s$^{-1}$ at room temperature (which increases to 1000 cm$^2$V$^{-1}$s$^{-1}$ at 120K, and higher at lower temperatures)[9] produces reasonably fast operations. These features are critical for building transistors with high current and power gains that are most important attributes for constructing high-frequency power amplifiers and high-speed logic circuits. The superiority of the phosphorene-based FETs has been experimentally demonstrated [9] (see Fig. 4a). In a vacuum environment (pressure ~1×10$^{-5}$ mbar), the channel switched from the 'on' state to the 'off' state and a drop in drain current by a factor of 10$^5$ (or 10$^4$)[10] were observed at room temperature, see Fig. 4b. The measured drain current modulation is four orders of magnitude larger than that in graphene and approaches the value recently reported in MoS$_2$–based devices. Subsequently, improved performances of phosphorene-FETs were achieved experimentally; for instance, the current fluctuations can be reduced with an Al$_2$O$_3$ overlayer protecting the device for better stability and reliability[92]. The high mobility in phosphorene yields fast operation, which can reach the gigahertz frequency level. Measurements showed a short-circuit current-gain cutoff frequency $f_T$ of 12 GHz and a maximum oscillation frequency $f_{max}$ of 20 GHz in 300 nm channel length devices. Devices based on phosphorene may offer advantages over graphene transistors for high frequency electronics in terms of voltage and power gain due to the good current saturation properties arising from their moderate band gap. Phosphorene thus can be considered as a promising candidate material for future high performance thin film electronics technology for operation in the multi-GHz frequency range and beyond.

With the appealing band gap and high mobility, phosphorene not only finds its promising application in transistors, but also exhibits outstanding performance in other semiconducting electronics, like circuits and amplitude modulated (AM) demodulators, which can act as an active stage for radio receivers and loudspeakers when connecting with audible signals[93]. As a proof-of-concept application, a flexible memory device using BP quantum dots mixed with polyvinylpyrrolidone as the active layer was successfully fabricated, and it exhibits a non-volatile rewritable memory effect with a high on/off current ratio and good stability.[94]

**Battery applications** Recent theoretical and experimental studies have indicated that phosphorus holds great promise for advanced battery applications, with a high theoretical specific capacity of 2596 mAh/g and a discharge potential range of 0.4-1.2 V. Theoretical calculations indicate[95] that Li atoms strongly bind with phosphorus atoms in the cationic state. The diffusion barrier is highly anisotropic, and its value along the zigzag direction is calculated to be 0.08 eV, leading to an ultrahigh diffusivity $10^2$ ($10^4$) times faster than that in $MoS_2$ (graphene) at room temperature. In contrast, the calculated barrier is at a much higher value of 0.68 eV along the armchair direction, effectively blocking diffusion in that direction. The predicted remarkably large average voltage of 2.9 V in a phosphorene-based Li-ion battery and good electrical conductivity of phosphorene as an electrode suggest great potential for its use as a rechargeable battery in portable electronics, electric vehicles, as well as large-scale stationary energy storage.

The application of bulk phosphorus as an anode is hindered by rapid capacity fading, which is induced by a large volume change (300%) upon lithiation and the consequent loss of electrical contact. Fabricated black phosphorus nanoparticle-graphite composites offer a possible solution[96]. The phosphorus-carbon bond formed in the composites is stable during the lithium insertion and extraction, which affords high capacity and cycle stability in the battery performance, thus maintaining excellent electrical connection between phosphorene and carbon materials. The discharge capacity of 2786 mAh/g achieved at a 0.2C current rate is significantly higher than that in lithium batteries with graphite, Ge and Sn as anodes, and comparable with that in the Si-anode battery. Its reliability is demonstrated with a good record of recycle life test, which shows that 80% of capacity is preserved after 100 cycles.

**Other directions of research and development** The application of phosphorene is of course not limited to transistors and batteries. Recent progress has led to concept demonstrations for optoelectronics[48] and photodetectors, which are capable of acquiring high-contrast images in visible and infrared spectral regime; photoluminescence quenching in phosphorene-TMDC heterostructures[50]; as well as photovoltaic effects when assembling with hexagonal BN as a p-n junction where a zero-bias photocurrent and significant open-circuit voltage can be obtained.[49] Ultrathin phosphorene-$MoS_2$ p-n diodes[52] have shown strong current-rectifying characteristics (Fig. 4c, 4d) and a maximum photo-detection responsivity of 418 mA/W upon illumination at the wavelength of 633 nm and photovoltaic energy conversion with an external quantum efficiency of 0.3%. However, it is fair to say that most of these devices are taking advantage of the appealing band gap (0.3-2 eV) and high mobility, such as in FET, AM demodulators and photodetectors.[9-10,47-52,92-97] The most unique feature of phosphorene - its in-plane anisotropy, which is intrinsic for its mechanical, electronic, electric, transport, thermoelectric, and optical properties – has not been widely used in device design. Explorations along this line could lead to the discovery of a gold mine for phosphorene applications and generate opportunities for designing conceptually new devices. One example is anisotropic nanomechanical resonators[98], which is suggested by numerical simulations for mechanical resonant responses of free standing phosphorene. The mechanical anisotropy operating in the elastic plate regime can lead to new multimode resonant characteristics in terms of mode sequences, shapes, and orientational preferences that are unavailable in nanomechanical resonators made of isotropic materials. The resonant response will depend on crystal orientation in asymmetric devices and the degree of anisotropy. They suggest a pathway towards harnessing the mechanical anisotropy in phosphorene for building novel 2D nanomechanical devices and resonant transducers with tunable multimode functions, but experimental verification is still required.

Thermoelectrics is another promising field of phosphorene application where the anisotropic transport properties have shown to remarkably improve the thermoelectric figure of merit, up to 2.5 [24], which has met the requirement of commercial use, but the experimental demonstration and the achievement of thermoelectric device are as yet still lacking. Another example is the potential application of phosphorene in a gas sensing device[30]. The superior gas sensor characteristic of phosphorene, with stronger binding energies compared with those of graphene and $MoS_2$, has already been confirmed from electronic and transport calculations. Again, the demonstration from experiments is still yet to come. It can be seen that, although a vibrant multiplicity of interesting research activities have been devoted to revealing fundamental properties and applications, the study for few-layer phosphorene is still in its infancy, with many unresolved issues and unexplored ideas.

## Challenges and possible solutions

Although BP, the parent material of phosphorene, is the most stable allotrope of elemental phosphorus and can be stable under ambient conditions for months, monolayer or few-layer phosphorene is found to be unstable in an atmospheric environment, being subject to severe degradation by moisture and oxygen in air[61], as well as molecular $N_2H_2$ [99]. Droplets due to water adsorption appear on the surface of BP flakes quickly after sample fabrication, which become visible on the surface and keep growing after one hour, see Fig. 5a and 5b. This strong hydrophilic character deteriorates BP, etching away the thinner parts of the flakes after a long (one week or more) exposure to air, where the volume will increase over 200% due to the condensation of moisture from air. The strong affinity of phosphorene for water greatly modifies the performance of the fabricated FETs measured in ambient conditions.[100] Upon exposure to air, in a short time (minutes), a shift in the threshold voltage occurs due to physisorbed oxygen and nitrogen. After longer exposure (hours), strong p-type doping occurs from water absorption. After continuous exposure in air, FETs eventually degrade and break-down after several days due to the layer-by-layer etching process.

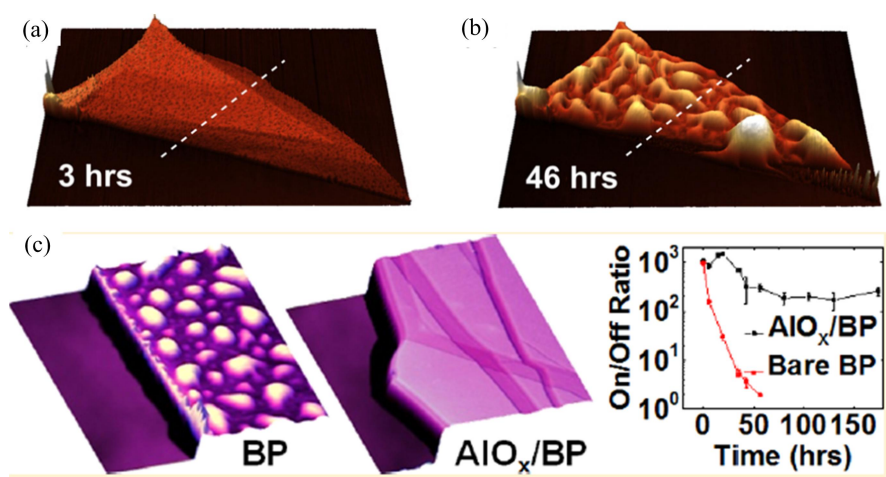

Figure 5. Degradation and encapsulation protection of phosphorene. (a) and (b) AFM scans of a phosphorene flake in air taken at 3 and 46 hours after exfoliation, respectively. (c) AFM images and On/Off ratio of BP FET without/with $AlO_x$ overlayer protection. **a,b: Reproduced from Ref. 100 with permission from IOP; c, Reprinted from Ref. 103.**

Theoretical predictions[61] indicate that water and phosphorene exhibit strong dipolar interactions. A large distortion of the lattice structure with around 25% of shrinking is found, which leads to a 20% modulation of the band gap. Aside from the physisorption-induced structural and electronic changes in phosphorene, chemisorption-induced oxidation is also a major cause for the degradation of phosphorene.[101] Oxygen in air easily forms a diagonal molecular bridge, with an energy gain of 0.13 eV with respect to isolated (triplet) oxygen and pristine phosphorene. Once the oxygen bridge has been formed, there is only a small (0.15 eV) energy cost for a phonon-mediated dissociation of the $O_2$ bridge, resulting in the formation of two dangling oxygen bonds, which significantly reduces the band gap. Depending on the oxygen concentration, oxidation of phosphorene can lead to the formation of a new family of planar (two-dimensional) and tubular (one-dimensional) oxides and suboxides, most of which are insulating with large band gaps up to 8.5 eV[102]. X-ray photoelectron spectroscopy characterization confirmed that $PO_x$ peaks appeared after exposure of BP to air for one day and the observed P-O stretching modes suggest that the formation of oxidized phosphorus leads to the property degradation[103].

Effective protection to prevent the degradation process is critical for phosphorene in practical applications. The device concept demonstrations in laboratories have employed high vacuum environments down to $10^{-5}$ mbar pressure to eliminate the agents (e.g., various gas molecules) that cause degradation[9], but this is unrealistic for large-scale implementation. Several alternative feasible methods have been developed such as encapsulation and coating; the effectiveness of this approach has been verified in silicene FET protection[104]. It also shows promise to protect BP flakes that can be stable for several weeks in ambient environments after being encapsulated by $Al_2O_3$ overlayers[92, 103]. In contrast, for unprotected BP FETs, the degradation causes large increases in the threshold voltage after 6 hours in the ambient environment, followed by a $\sim 10^3$ decrease in FET current on/off ratio and mobility after 48 hours, see Fig. 5c. The deposited $AlO_x$ overlayers effectively suppress ambient degradation, allowing encapsulated BP FETs to maintain high on/off ratios of $\sim 10^3$ and mobilities of $\sim 100$ cm$^2$V$^{-1}$s$^{-1}$ for over 2 weeks in ambient conditions[103]. Encapsulated BP/phosphorene on flexible polyimide, graphene, or BN have also been proposed for the same reasons with different levels of success.[105] Although phosphorene is clearly not as stable as graphene and $MoS_2$, commercial and laboratory successes in protection of unstable materials such as silicene suggest that the development of passivation, encapsulation and packaging technologies, which could protect the relatively vulnerable 2D material from the environmental attack, may provide resolutions to these issues.

## Summary and outlook

As a new member of the 2D layered materials family, phosphorene displays many outstanding properties that have advantageous over those of other prominent layered materials like graphene and TMDCs; of particular significance is that phosphorene combines high charge mobility and desirable band gap that are indispensable for the recently demonstrated excellent performance of phosphorene-based devices such as the FETs. The layer-stacked configuration of bulk BP allows relatively easy fabrication of monolayer or few-layer phosphorene via mechanical exfoliation, analogous to the procedure that produces graphene from graphite. But the absence of chemical growth for monolayer and few-layer phosphorene limits the quality and size of fabricated samples, and the development along the direction based on the successful CVD growth of graphene, silicene and TMDC layers is promising to accelerate its extensive applications in nanodevice design and implementation.

Extensive theoretical and experimental investigations have revealed novel electronic, mechanical, and transport properties that exhibit intrinsic anisotropic behaviour. This unique characteristic makes phosphorene outstanding amongst the burgeoning family of 2D layered materials with great promise for its applications in electronics, spintronics and photovoltaics. Most of the current researches focus on applications and device demonstrations of phosphorene based on its appealing gap sizes and high electron mobility; however, its most unique features, namely the highly anisotropic structural, electronic and transport properties, have been largely unexplored, and this points to new opportunities for future research and development activities. Further work in this direction may reveal more superior performance characteristics of phosphorene over those available in devices based on graphene and TMDCs. However, as always, opportunities and challenges coexist in the study and utilization of new materials, and such is the case for phosphorene. Obstructions such as degradation effects caused by oxidation and hydrophilicity must be overcome before practical applications can ensue, and effective strategies are emerging to reduce or even largely eliminate the structural and property degradation using the encapsulation and surface passivation techniques.

Overall, many of the phosphorene's intrinsically superior properties make it a promising material for nanodevice design and fabrication, and more fundamental and technological breakthroughs can be expected in the near future.

## Acknowledgements

SCS gratefully acknowledges support from the UNSW SPF01 scheme. CFC was supported in part by the Department of Energy through the Cooperative Agreement DE-NA0001982.

## Notes

The authors declare no competing financial interest.

## Biographies

**Dr. Liangzhi Kou** received his Ph.D. in 2011 from Nanjing University of Aeronautics and Astronautics. After two years as an Alexander von Humboldt fellow at the Bremen Center of Computational Materials Sciences (BCCMS), he joined the Integrated Materials Design Centre (IMDC) at UNSW Australia in 2014. He mainly focuses on computational discovery and design of novel 2D materials for energy applications and 2D topological insulators.

**Prof. Changfeng Chen** received his Ph.D. from Peking University in 1987. After postdoctoral appointments at UC Berkeley and University of Oregon, he joined University of Nevada, Las Vegas, where he is now a Professor of Physics. His research interests are in the areas of materials under extreme pressure and temperature, new structural forms and novel phenomena in nanoscale materials, and strongly correlated electron systems.

**Prof. Sean Smith** received his Ph.D. from the University of Canterbury, New Zealand, in 1989. After a Humboldt fellowship at Göttingen University and postdoctoral research at UC Berkeley, he held faculty appointments at The University of Queensland (1993-2011) and was a Center Director and Chief Scientist at Oak Ridge National Laboratory (ORNL, 2011-2014) before moving to UNSW Australia in 2014, where he directs the IMDC. He currently explores structure, complexation, kinetics, dynamics, and catalysis within nanomaterials and hybrid nanobio systems.

## References

(1)   Chou, T.-D.; Lee, T.-W.; Chen, S.-L.; Tung, Y.-M.; Dai, N.-T.; Chen, S.-G.; Lee, C.-H.; Chen, T.-M.; Wang, H.-J. The Management of White Phosphorus Burns, *Burns* **2001**, *27*, 492–497

(2)   Butler, S. Z.; Hollen, S. M.; Cao, L.; Cui, Y.; Gupta, J. A.; Gutiérrez, H. R.; T. Heinz, F.; Hong,
      S. S.; Huang, J.; Ismach, A. F.; et. al. Progress, Challenges, and Opportunities in Two-
      Dimensional Materials Beyond Graphene, *ACS Nano*, **2013**, *7*, 2898-2926.

(3)   Geim, A. K. and Novoselov, K. S. The Rise of Graphene, *Nat. Mater.*, **2007**, *6*, 183–191.

(4)   Wang, H.; Yuan, H.; Hong, S. S.; Li, Y. and Cui, Y. Physical and Chemical Tuning of Two-
      dimensional Transition Metal Dichalcogenides, *Chem. Soc. Rev.*, **2015**, *44*, 2664-2680

(5)   Wang, Q. H.; Kalantar-Zadeh, K.; Kis, A.; Coleman, J. N. and Strano, M. S. Electronics and
      Optoelectronics of Two-dimensional Transition Metal Dichalcogenides, *Nat. Nanotech.* **2012**, *7*,
      699–712.

(6)   Appalakondaiah, S.; Vaitheeswaran, G.; Lebègue, S.; Christensen, N. E. and Svane, A. Effect of
      van der Waals Interactions on the Structural and Elastic Properties of Black Phosphorus, *Phys.
      Rev. B* **2012**, *86*, 035105.

(7)   Hultgren, R.; Gingrich, N. S. and Warren, B. E. The Atomic Distribution in Red and Black
      Phosphorus and the Crystal Structure of Black Phosphorus, *J. Chem. Phys.* **1935**, *3*, 351-355.

(8)   Thurn, H. and Kerbs, H. Crystal Structure of Violet Phosphorus, *Angew. Chem., Int. Ed.* **1966**, *5*,
      1047-1048.

(9)   Li, L.; Yu, Y.; Ye, G. J.; Ge, Q.; Ou, X.; Wu, H.; Feng, D.; Chen, X. H. and Zhang, Y. Black
      Phosphorus Field-effect Transistors. *Nat. Nanotech.* **2014**, *9*, 372–377.

(10)  Liu, H.; Neal, A. T.; Zhu, Z.; Luo, Z.; Xu, X.; Tománek, D. and Ye, P. D. Phosphorene: An
      Unexplored 2D Semiconductor with a High Hole Mobility. *ACS Nano*, **2014**, *8*, 4033–4041.

(11)  Ling, X.; Wang, H.; Huang, S.; Xia, F. and Dresselhaus, M. S. The Renaissance of Black
      Phosphorus, *Proc. Natl. Acad. Sci. USA*, **2015,** *112*, 4523–4530

(12)  Liu, H.; Du, Y.; Deng, Y. and Ye, P. D. Semiconducting Black Phosphorus: Synthesis,
      Transport Properties and Electronic Applications, *Chem. Soc. Rev.*, **2015**, *44*, 2732-2743

(13)  Dai, J. and Zeng, X. C. Bilayer Phosphorene: Effect of Stacking Order on Bandgap and Its
      Potential Applications in Thin-Film Solar Cells, *J. Phys. Chem. Lett.*, **2014**, *5*, 1289-1293.

(14)  Novoselov, K. S.; Jiang, D.; Schedin, F.; Booth, T. J.; Khotkevich, V. V.; Morozov, S. V. and
      Geim, A. K. Two-dimensional Atomic Crystals, *Proc. Natl. Acad. Sci. USA*, **2005**, *102*, 10451-
      10453.

(15)  Lee, C.; Yan, H.; Brus, L. E.; Heinz, T. F.; Hone, J. and Ryu, S. Anomalous Lattice Vibrations
      of Single- and Few-layer $MoS_2$. *ACS Nano*, **2010**, *4*, 2695–2700.

(16)  Mak, K. F.; Lee, C.; Hone, J.; Shan, J. and Heinz, T. F. Atomically Thin $MoS_2$: a New Direct-
      Gap Semiconductor. *Phys. Rev. Lett.* **2010**, *105*, 136805.

(17)  Liu, C.-C.; Feng, W. and Yao, Y. Quantum Spin Hall Effect in Silicene and Two-Dimensional
      Germanium, *Phys. Rev. Lett.* **2011**, *107*, 076802.

(18)  Rodin, A. S.; Carvalho, A. and Castro Neto, A. H. Strain-Induced Gap Modification in Black
      Phosphorus, *Phys. Rev. Lett.* **2014**, *112*, 176801.

(19) Fei, R. and Yang, L. Strain-Engineering the Anisotropic Electrical Conductance of Few-Layer Black Phosphorus, *Nano Lett*. **2014**, *14*, 2884−2889

(20) Xu, Y.; Dai, J. and Zeng, X. C. Electron-Transport Properties of Few-Layer Black Phosphorus, *J. Phys. Chem. Lett*., **2015**, *6*, 1996-2002

(21) Qin, G.; Yan, Q.-B.; Qin, Z.; Yue, S.-Y.; Cui, H.-J. Zheng, Q.-R. and Su, G. Hinge-like Structure Induced Unusual Properties of Black Phosphorus and New strategies to Improve the Thermoelectric Performance, *Sci. Rep*. **2014**, *4*, 6946.

(22) Ong, Z.-Y.; Cai, Y.; Zhang, G. and Zhang, Y.-W. Strong Thermal Transport Anisotropy and Strain Modulation in Single-Layer Phosphorene, *J. Phys. Chem. C* **2014**, *118*, 25272−25277

(23) Jain, A. and McGaughey, A. J. H. Strongly Anisotropic In-plane Thermal Transport in Single-layer Black Phosphorene, *Sci. Rep*. **2014**, *5*, 8501.

(24) Fei, R.; Faghaninia, A.; Soklaski, R.; Yan, J.-A.; Lo, C. and Yang, L. Enhanced Thermoelectric Efficiency via Orthogonal Electrical and Thermal Conductances in Phosphorene, *Nano Lett*. **2014**, *14*, 6393−6399.

(25) Wei, Q. and Peng, X. Superior Mechanical Flexibility of Phosphorene and Few-layer Black Phosphorus, *Appl. Phys. Lett*. **2014**, *104*, 251915.

(26) Wang, L.; Kutana, A.; Zou, X. and Yakobson, B. I. Electro-Mechanical Anisotropy of Phosphorene, *Nanoscale*, **2015**, *7*, 9746-9751.

(27) Jiang, J.-W.; Park, H. S. Negative Poisson's Ratio in Single-Layer Black Phosphorus *Nat. Comm*. **2014**, *5*, 4727

(28) Jiang, J.-W.; Park, H. S. Mechanical Properties of Single-Layer Black Phosphorus, *J. Phys. D: Appl. Phys*. **2014**, *47*, 385304

(29) Low, T.; Roldán, R.; Wang, H.; Xia, F.; Avouris, P.; Moreno, L. M. and Guinea, F. Plasmons and Screening in Monolayer and Multilayer Black Phosphorus, *Phys. Rev. Lett*. **2014**, *113*, 106802.

(30) Kou, L.; Frauenheim, T.; Chen, C. Phosphorene as a Superior Gas Sensor: Selective Adsorption and Distinct I–V Response, *J. Phys. Chem. Lett*. **2014**, *5*, 2675-2681

(31) Kou, L.; Ma. Y.; Smith. S. C.; Chen, C. Anisotropic Ripple Deformation in Phosphorene, *J. Phys. Chem. Lett.,* **2015**, *6*, 1509-1513

(32) Zhang, R.; Li, B. and Yang, J.; A First-Principles Study on Electron Donor and Acceptor Moluecles Adsorbed on Phosphorene, *J. Phys. Chem. C*., **2015**, *119*, 2871-2878.

(33) Sui, X.; Si, C.; Shao, B.; Zou, X.; Wu, J.; Gu, B.-L.; and Duan, W. Tunable Magnetism in Transition-Metal-Decorated Phosphorene, *J. Phys. Chem. C*., **2015**, *119*, 10059–10063.

(34) Jing, Y.; Tang, Q.; He, P.; Zhou, Z.; and Shen, P.; Small Molecules Make Big Differences: Molecular Doping Effects on Electronic and Optical Properties of Phosphorene, *Nanotechnology*, **2015**, *26*, 095201.

(35) Hu, T. and Hong, J. First-Principles Study of Metal Adatom Adsorption on Black Phosphorene,

*J. Phys. Chem. C*, **2015**, *119*, 8199–8207

(36) Guo, H.; Lu, N.; Dai, J.; Wu, X. and Zeng, X. C. Phosphorene Nanoribbons, Phosphorus Nanotubes, and van der Waals Multilayers. *J. Phys. Chem. C*, **2014**, *118*, 14051–14059

(37) Han, X.; Stewart, H. M.; Shevlin, S. A.; Catlow, C. R. A.; Guo, Z. X. Strain and Orientation Modulated Bandgaps and Effective Masses of Phosphorene Nanoribbons. *Nano Lett***. 2014,** *14*, 4607–4614.

(38) Tran, V.; Yang, L.; Scaling Laws for the Band Gap and Optical Response of Phosphorene Nanoribbons. *Phys Rev B* **2014,** *89*, 245407

(39) Peng, X.; Copple, A.; Wei, Q.; Edge Effects on the Electronic Properties of Phosphorene Nanoribbons. *J. Appl. Phys*. **2014**, *116*, 144301

(40) Liang, L.; Wang, J.; Lin, W.; Sumpter, B. G.; Meunier, V. and Pan, M. Electronic Bandgap and Edge Reconstruction in Phosphorene Materials, *Nano Lett*. **2014**, *14*, 6400–6406.

(41) Guan, J.; Zhu, Z. and Tománek, D. Phase Coexistence and Metal-Insulator Transition in Few-Layer Phosphorene: A Computational Study, *Phys. Rev. Lett*. **2014**, *113*, 046804.

(42) Zhu, Z. and Tománek, D. Semiconducting Layered Blue Phosphorus: A Computational Study, *Phys. Rev. Lett*. **2014**, *112*, 176802.

(43) Wu, M.; Fu, H.; Zhou, L.; Yao, K. and Zeng, X. C. Nine New Phosphorene Polymorphs with Non-Honeycomb Structures: A Much Extended Family, *Nano Lett*. **2015**, *15*, 3557-3562.

(44) Liu, Q.; Zhang, X.; Abdalla, L. B.; Fazzio, A. and Zunger, A. Switching a Normal Insulator into a Topological Insulator via Electric Field with Application to Phosphorene, *Nano Lett*. **2015**, *15*, 1222−1228

(45) Ge, Y.; Wan, W.; Yang, F. and Yao, Y. The Strain Effect on Superconductivity in Phosphorene: A First Principles Prediction, *New J. Phys*. **2015**, *17*, 035008

(46) L. Li, G. J. Ye, V. Tran, R. Fei, G. Chen, H. Wang, J. Wang, K. Watanabe, T. Taniguchi, L. Yang, X. H. Chen and Y. Zhang, Quantum Oscillations in a Two-dimensional Electron Gas in Black Phosphorus Thin Films, *Nat. Nanotehnol*., **2015**, doi:10.1038/nnano.2015.91.

(47) Das, S.; Demarteau, M. and Roelofs, A. Ambipolar Phosphorene Field Effect Transistor, *ACS Nano*, **2014**, *8*, 11730–11738

(48) Xia, F.; Wang, H.; Jia, Y.; Rediscovering Black Phosphorus as an Anisotropic Layered Material for Optoelectronics and Electronics. *Nat. Comm*. **2014**, *5*, 4458.

(49) Buscema, M.; Groenendijk, D. J.; Steele, G.; A.; van der Zant, H. S. J.; Castellanos-Gomez, A.; Photovoltaic Effect in Few-layer Black Phosphorus PN Junctions Defined by Local Electrostatic Gating. *Nat. Comm*. **2014**, *5*, 4651.

(50) Hong, T.; Chamlagain, B.; Lin, W.; Chuang, H.-J.; Pan, M.; Zhou, Z. and Xu, Y.-Q. Polarized Photocurrent Response in Black Phosphorus Field-effect Transistors. *Nanoscale*, **2014**, *6*, 8978–8983

(51) Yuan, J.; Najmaei, S.; Zhang, Z.; Zhang, J.; Lei, S.; Ajayan, P. M.; Yakobson, B. I. and Lou, J.

Photoluminescence Quenching and Charge Transfer in Artificial Heterostacks of Monolayer Transition Metal Dichalcogenides and Few-Layer Black Phosphorus, *ACS Nano*, **2015**, *9*, 555-563.

(52) Deng, Y.; Luo, Z.; Conrad, N. J.; Liu, H.; Gong, Y.; Najmaei, S.; Ajayan, P. M.; Lou, J.; Xu, X. and Ye, P. D. Black Phosphorus Monolayer $MoS_2$ van der Waals Heterojunction p-n Diode, *ACS Nano*, **2014**, *8*, 8292-8299.

(53) Cai, Y.; Zhang, G. and Zhang, Y.-W.; Electronic Properties of Phosphorene/Graphene and Phosphorene/Hexagonal Boron Nitride Heterostructures, *J. Phys. Chem. C*, **2015**, DOI: 10.1021/acs.jpcc.5b02634.

(54) Cunningham, G.; Lotya, M.; Cucinotta, C. S.; Sanvito, S.; Bergin, S. D.; Menzel, R.; Shaffer, M. S. P. and Coleman, J. N.; Solvent Exfoliation of Transition Metal Dichalcogenides: Dispersibility of Exfoliated Nanosheets Varies Only Weakly between Compounds. *ACS Nano*, **2012**, *6*, 3468–3480.

(55) Díaz, E.; Ordóñez, S. and Vega, A. Adsorption of Volatile Organic Compounds onto Carbon Nanotubes, Carbon Nanofibers, and High-surface-area Graphites. *J. Colloid Interface Sci.* **2007**, *305*, 7-16.

(56) Hernandez, Y.; Nicolosi, V.; Lotya, M.; Blighe, F. M.; Sun, Z.; De, S.; McGovern, I. T.; Holland, B.; Byrne, M.; Gun'Ko, Y. K.; et al. High-yield Production of Graphene by Liquid-phase Exfoliation of Graphite. *Nat. Nanotech.* **2008**, *3*, 563–568.

(57) Alem, N.; Erni, R.; Kisielowski, C.; Rossell, M. D.; Gannett, W. and Zettl, A. Atomically Thin Hexagonal Boron Nitride Probed by Ultrahigh Resolution Transmission Electron Microscopy. *Phys. Rev. B* **2009**, *80*, 155425.

(58) Yang, X.; Tang, S.; Ding, G.; Xie, X.; Jiang, M. and Huang, F. Layer-by-layer Thinning of Graphene by Plasma Irradiation and Post-annealing, *Nanotechnology*, **2012**, *23*, 025704

(59) Liu, Y. L.; Nan, H.; Wu, X.; Pan, W.; Wang, W.; Bai, J.; Zhao, W.; Sun, L.; Wang, X. and Ni. Z. Layer-by-layer Thinning of $MoS_2$ by Plasma. *ACS Nano*, **2013**, *7*, 4202–4209.

(60) Lu, W.; Nan, H.; Hong, J.; Chen, Y.; Zhu, C.; Liang, Z.; Ma, X.; Ni, Z.; Jin, C. and Zhang, Z. Plasma-assisted Fabrication of Monolayer Phosphorene and Its Raman Characterization, *Nano Res.*, **2014**, *7*, 853–859.

(61) Castellanos-Gomez, A.; Vicarelli, L.; Prada, E.; Island, J. O.; Narasimha-Acharya, K. L.; Blanter, S. I.; Groenendijk, D. J.; Buscema, M.; Steele, G. A.; Alvarez, J. V.; et. al. Isolation and Characterization of Few-layer Black Phosphorus, *2D Mater.*, **2014**, *1*, 025001.

(62) V. Nicolosi, M. Chhowalla, M. G. Kanatzidis, M. S. Strano, J. N. Coleman, Liquid Exfoliation of Layered Materials, *Science*, **2013**, *340*, 1226419

(63) Coleman, J. N.; Lotya, M.; O'Neill, A.; Bergin, S. D.; King, P. J.; Khan, U.; Young, K.; Gaucher, A.; De, S.; Smith, R. J.; et. al. Two-Dimensional Nanosheets Produced by Liquid Exfoliation of

Layered Materials, *Science* **2011**, *331*, 568-571

(64) Yasaei, P.; Kumar, B.; Foroozan, T.; Wang, C.; Asadi, M.; Tuschel, D.; Indacochea, J. E.; Klie, R. F. and Salehi-Khojin, A. High-Quality Black Phosphorus Atomic Layers by Liquid-Phase Exfoliation, *Adv. Mater.* **2015**, *27*, 1887–1892

(65) Brent, J. R.; Savjani, N.; Lewis, E. A.; Haigh, S. J.; Lewis, D. J.; O'Brien, P. Production of Few-layer Phosphorene by Liquid Exfoliation of Black Phosphorus, *Chem. Comm.* **2014** , *50*, 13338-13341.

(66) Zeng, Z.; Sun, T.; Zhu, J.; Huang, X.; Yin, Z.; Lu, G.; Fan, Z.; Yan, Q.; Hng, H. H. and Zhang, H. An Effective Method for the Fabrication of Few-layer-thick Inorganic Nanosheets. *Angew. Chem. Int. Ed.* **2012**, *51*, 9052–9056.

(67) Zeng, Z. Y.; Yin, Z.; Huang, X.; Li, H.; He, Q.; Lu, G.; Boey, F.; and Zhang, H. Single-layer Semiconducting Nanosheets: High-yield Preparation and Device Fabrication. *Angew. Chem. Int. Ed.* **2011**, *50*, 11093-11097.

(68) Zhao, S.; Kang, W. and Xue, J. The Potential Application of Phosphorene as an Anode Material in Li-ion Batteries, *J. Mater. Chem. A*, **2014**, *2*, 19046-19052

(69) Li, X.; Cai, W.; An, J.; Kim, S.; Nah, J.; Yang, D.; Piner, R.; Velamakanni, A.; Jung, I.; Tutuc, E.; et. al. Large-area Synthesis of High-quality and Uniform Graphene Films on Copper Foils. *Science* **2009**, *324*, 1312-1314.

(70) Lee, Y.-H.; Zhang, X.-Q.; Zhang, W.; Chang, M.-T.; Lin, C.-T.; Chang, K.-D.; Yu, Y.-C.; Wang, J. T.-W.; Chang, C.-S.; Li, L.-J. and Lin, T.-W.; Synthesis of Large-area $MoS_2$ Atomic Layers with Chemical Vapor Deposition. *Adv. Mater.* **2012**, *24*, 2320–2325.

(71) Liu, K.-K.; Zhang, W.; Lee, Y.-H.; Lin, Y.-C.; Chang, M.-T.; Su, C.-Y.; Chang, C.-S.; Li, H.; Shi, Y.; Zhang, H.; et al. Growth of Large-area and Highly Crystalline $MoS_2$ Thin Layers on Insulating Substrates. *Nano Lett.* **2012**, *12*, 1538–1544.

(72) Hass, J.; de Heer, W. A. and Conrad, E. H. The Growth and Morphology of Epitaxial Multilayer Graphene. *J. Phys. Condens. Matter* **2008**, *20*, 323202.

(73) Köpf, M.; Eckstein, N.; Pfister, D.; Grotz, C.; Krüger, I.; Greiwe, M.; Hansen, T.; Kohlmann, H.; Nilges, T. Access and in Situ Growth of Phosphorene-precursor Black Phosphorus, *J. Cryst. Growth* **2014,** *405*, 6–10.

(74) Lalmi, B.; Oughaddou, H.; Enriquez, H.; Kara, A.; Vizzini, S.; Ealet, B. and Aufray, B. Epitaxial Growth of A Silicene Sheet, *Appl. Phys. Lett.* **2010**, *97*, 223109.

(75) Chen, P.; Yang, J.-J.; Li, S.-S.; Wang, Z.; Xiao, T.; Qian, Y.-H.; Yun, S.-H.; Hydrothermal Synthesis of Macroscopic Nitrogen-doped Graphene Hydrogels for Ultrafast Supercapacitor, *Nano Energy* **2013,** *2*, 249–256

(76) Peng, Y.; Meng, Z.; Zhong, C.; Lu, J.; Yu, W.; Yang, Z.; Qian, Y.; Hydrothermal Synthesis of $MoS_2$ and Its Pressure-related Crystallization. *J. Solid. State. Chem.* **2001**, *159*, 170–173.

(77) Tran, V.; Soklaski, R.; Liang, Y.; Yang L.; Layer-controlled Band Gap and Anisotropic Excitons

in Few-layer Black Phosphorus. *Phys. Rev. B*, **2014**, *89*, 235319.

(78)  Fuhrer, M. S. and Hone, J.; Measurement of Mobility in Dual-gated MoS$_2$ Transistors, *Nat. Nanotechnol*. **2013**, *8*, 146-147.

(79)  Radisavljevic, B. and Kis, A.; Reply to 'Measurement of Mobility in Dual-gated MoS$_2$ Transistors', *Nat. Nanotechnol*. **2013**, *8*, 147-148.

(80)  Peng, X.; Wei, Q.; Copple A.; Strain-engineered Direct-indirect Band Gap Transition and Its Mechanism in Two-dimensional Phosphorene. *Phys Rev B* **2014**, *90*, 085402.

(81)  Manjanath, A.; Samanta, A.; Pandey, T. and Singh, A. K., Semiconductor to Metal Transition in Bilayer Phosphorene under Normal Compressive Dtrain, *Nanotechnology*, **2015**, *26*, 075701

(82)  Dragoman, D.; Dragoman, M. Giant Thermoelectric Effect in Graphene, *Appl. Phys. Lett*., **2007**, *91*, 203116.

(83)  Balandin, A. A. Thermal Properties of Graphene and Nanostructured Carbon Materials, *Nat. Mater*. **2011**, *10*, 569-581.

(84)  Sahoo, S.; Gaur, A. P. S.; Ahmadi, M.; Guinel, M. J.-F. and Katiyar, R. S. Temperature-Dependent Raman Studies and Thermal Conductivity of Few-Layer MoS$_2$, *J. Phys. Chem. C* **2013**, *117*, 9042−9047

(85)  Huang, W.; Da, H. and Liang, G.; Thermoelectric Performance of MX$_2$ (M = Mo, W; X = S, Se) Monolayers, *J. Appl. Phys*. **2013**, *113*, 104304.

(86)  Peelaers, H. and Van de Walle, C. G. Effects of Strain on Band Structure and Effective Masses in MoS$_2$, *Phys. Rev. B* **2012**, *86*, 241401(R).

(87)  Bolotin, K. I.; Sikes, K. J.; Jiang, Z.; Klima, M.; Fudenberg, G.; Hone, J.; Kim, P.; Stormer, H. L. Ultrahigh Electron Mobility in Suspended Graphene, *Solid State Comm*. **2008**, *146*, 351-355.

(88)  Das, A.; Pisana, S.; Chakraborty, B.; Piscanec, S.; Saha, S. K.; Waghmare, U. V.; Novoselov, K. S.; Krishnamurthy, H. R.; Geim, A. K.; Ferrari, A. C. and Sood, A. K.; Monitoring Dopants by Raman Scattering in an Electrochemically Top-gated Graphene Transistor, *Nat. Nanotechnol*, **2008**, *3*, 210-215.

(89)  Liu, F.; Ming, P. and Li, J. Ab Initio Calculation of Ideal Strength and Phonon Instability of Graphene under Tension, *Phys. Rev. B* **2007**, *76*, 064120.

(90)  Bertolazzi, S.; Brivio, J. and Kis, A. Stretching and Breaking of Ultrathin MoS$_2$, *ACS Nano* **2011**, *5*, 9703-9709.

(91)  Li, T. Ideal Strength and Phonon Instability in Single-layer MoS$_2$, *Phys. Rev. B* **2012**, 85, 235407.

(92)  Na, J.; Lee, Y. T.; Lim, J. A.; Hwang, D. K.; Kim, G.-T.; Choi, W. K. and Song, Y.-W. Few-layer Black Phosphorus Field-effect Transistors with Reduced Current Fluctuation. *ACS Nano* **2014**, *8*, 11753−11762.

(93)  Zhu, W.; Yogeesh, M. N.; Yang, S.; Aldave, S. H.; Kim, J.-S.; Sonde, S.; Tao, L.; Lu, N. and Akinwande, D. Flexible Black Phosphorus Ambipolar Transistors, Circuits and AM

Demodulator, *Nano Lett.*, **2015**, *15*, 1883–1890

(94) Zhang, X.; Xie, H.; Liu, Z.; Tan, C.; Luo, Z.; Li, H.; Lin, J.; Sun, L.; Chen, W.; Xu, Z.; et. al. Black Phosphorus Quantum Dots, *Angew. Chem. Int. Ed.* **2015**, *54*, 3724–3728

(95) Li, W.; Yang, Y.; Zhang, G. and Zhang, Y.-W. Ultrafast and Directional Diffusion of Lithium in Phosphorene for High-Performance Lithium-Ion Battery, *Nano Lett.*, **2015**, *15*, 1691–1697

(96) Sun, J.; Zheng, G.; Lee, H.-W.; Liu, N.; Wang, H.; Yao, H.; Yang, W. and Cui, Y. Formation of Stable Phosphorus−Carbon Bond for Enhanced Performance in Black Phosphorus Nanoparticle−Graphite Composite Battery Anodes, *Nano Lett.* **2014**, *14*, 4573−4580

(97) Engel, M.; Steiner, M.; Avouris, P. Black Phosphorus Photodetector for Multispectral, High-Resolution Imaging, *Nano Lett.* **2014**, *14*, 6414– 6417

(98) Wang, Z. and Feng, P. X.-L. Design of Black Phosphorus 2D Nanomechanical Resonators by Exploiting the Intrinsic Mechanical Anisotropy, *2D Mater.* **2015**, *2*, 021001.

(99) Dai, J. and Zeng, X. C.; Structure and stability of two dimensional phosphorene with =O or =NH functionalization, *RSC Adv.* **2014**, *4*, 48017-48021.

(100) Island, J. O.; Steele, G. A.; van der Zant, H. S. J. and Castellanos-Gomez, A. Environmental Instability of Few-layer Black Phosphorus, *2D Mater*. **2015**, *2*, 011002.

(101) Ziletti, A.; Carvalho, A.; Campbell, D. K.; Coker, D. F. and Castro Neto, A. H. Oxygen Defects in Phosphorene, *Phys. Rev. Lett.* **2015**, 114, 046801.

(102) Ziletti, A.; Carvalho, A.; Trevisanutto, P. E.; Campbell, D. K.; Coker, D. F. and Castro Neto, A. H. Phosphorene Oxides: Bandgap Engineering of Phosphorene by Oxidation, *Phys. Rev. B* **2015**, *91*, 085407.

(103) Wood, J. D.; Wells, S. A.; Jariwala, D.; Chen, K.-S.; Cho, E.; Sangwan, V. K.; Liu, X.; Lauhon, L. J.; Marks, T. J. and Hersam, M. C.; Effective Passivation of Exfoliated Black Phosphorus Transistors against Ambient Degradation. *Nano Lett* **2014**, *14*, 6964–6970.

(104) Tao, L.; Cinquanta, E.; Chiappe, D.; Grazianetti, C.; Fanciulli, M.; Dubey, M.; Molle, A. and Akinwande, D., Silicene Field-effect Transistors Operating at Room Temperature, *Nature Nanotech.* **2015**, *10*, 227-231.

(105) Avsar, A.; Vera-Marun, I. J.; Tan, J. Y.; Watanabe, K.; Taniguchi, T.; Castro Neto, A. H. and Ozyilmaz, B. Air-Stable Transport in Graphene-Contacted, Fully Encapsulated Ultrathin Black Phosphorus-Based Field-Effect Transistors, *ACS Nano*, **2015**, *9*, 4138–4145

**Quotes to highlight in paper**

1. Advanced chemical synthesis of phosphorene may draw inspiration from the techniques used for graphene and TMDCs, such as hydrothermal synthesis or CVD growth.

2. Phosphorene exhibits a comprehensive range of desirable physical parameters for device applications, making it a more favorable choice on balance than graphene and $MoS_2$.

3. The most unique feature of phosphorene - its in-plane anisotropy, which is intrinsic for its mechanical, electronic, electric, transport, thermoelectric, and optical properties – has not been widely used in device design.

4. Although phosphorene is clearly not as stable as graphene and $MoS_2$, commercial and laboratory successes in protection of unstable materials such as silicene suggest that the development of passivation, encapsulation and packaging technologies, which could protect the relatively vulnerable 2D material from the environmental attack, may provide resolutions to these issues.